# Design, Manufacture and Initial Operation of the Beryllium Components of the JET ITER-Like Wall

V Riccardo[a], P Lomas[a], G F Matthews[a], I Nunes[b], V Thompson[1a], E Villedieu[c]
and JET EFDA Contributors*

*JET-EFDA, Culham Science Centre, Abingdon, OX14 3DB, UK*
[a]*EURATOM/CCFE Fusion Association, Culham Science Centre, Abingdon, OX14 3DB, UK*
[b]*Associação EURATOM-IST, IPFN - Laboratório Associado, IST, Lisboa, Portugal*
[c]*CEA, IRFM, F-13108 Saint-Paul-lez-Durance, France*
\* See the Appendix of F. Romanelli et al., Proceedings of the 23rd IAEA Fusion Energy Conference 2010, Daejeon, Korea

The aim of the JET ITER-like Wall Project was to provide JET with the plasma facing material combination now selected for the DT phase of ITER (bulk beryllium main chamber limiters and a full tungsten divertor) and, in conjunction with the upgraded neutral beam heating system, to achieve ITER relevant conditions.

The design of the bulk Be plasma facing components had to be compatible with increased heating power and pulse length, as well as to reuse the existing tile supports originally designed to cope with disruption loads from carbon based tiles and be installed by remote handling. Risk reduction measures (prototypes, jigs, etc) were implemented to maximize efficiency during the shutdown. However, a large number of clashes with existing components not fully captured by the configuration model occurred.

Restarting the plasma on the ITER-like Wall proved much easier than for the carbon wall and no deconditioning by disruptions was observed. Disruptions have been more threatening than expected due to the reduced radiative losses compared to carbon, leaving most of the plasma magnetic energy to be conducted to the wall and requiring routine disruption mitigation. The main chamber power handling has achieved and possibly exceeded the design targets.

## 1. Introduction

The aim of the JET ITER-like Wall (ILW) Project was to provide JET with the plasma facing material combination now selected for the DT phase of ITER (bulk Be main chamber limiters and a W divertor) and, in conjunction with the upgraded neutral beam heating system, to achieve ITER relevant conditions [1].

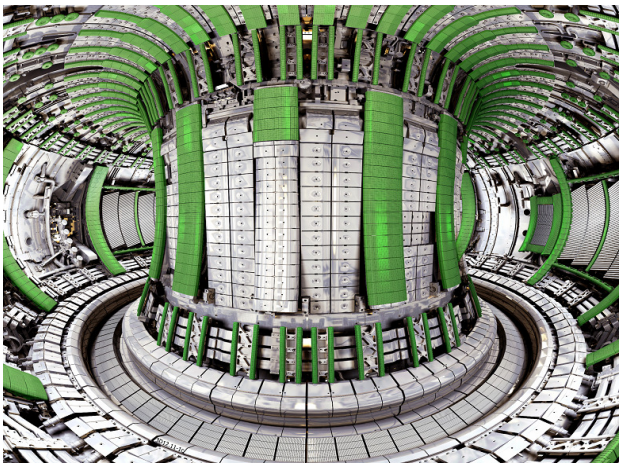

Fig.1 Be tile assemblies highlighted (in green on-line only)

The new Be plasma facing surface represents ~40m$^2$ of the ~80m$^2$ replaced by the ILW, the regions covered by the Be tile assemblies are shown in Fig. 1. Section 2 and 3 describe respectively the design and the manufacture of the ILW Be plasma facing components (PFCs). Section 4 covers the installation of the Be PFCs as well as other issues not driven by their design and manufacture, but that were addressed by modifying the ILW Be PFCs. Finally section 5 discusses engineering aspects of the operation with the ILW Be PFCs.

## 2. Design of ILW Be PFCs

The design had to be compatible with increased heating power and pulse length, from 24 MW to 35 MW and from 10 s to 20 s respectively, as part of the neutral beam upgrade carried out in parallel. The constraints were to reuse the existing tile supports (originally designed for disruption loads from C-based tiles and providing no active cooling) and to be installed by remote handling (RH). As a result, the Be tile assemblies were sized to fit within the RH payload while offering the maximum possible thermal capacity. Their sliced design is a compromise between minimizing eddy loads and optimizing the power handling. The latter was further improved by avoiding plasma facing bolting holes, resulting in sequential beam assembly.

### 2.1 Compatibility with disruption loads

Plasma disruptions produce changes in magnetic field of the order of 100 T/s, which induce eddy currents in the conducting materials. The C-based tiles have electrical resistivity of ~10μΩm and eddy currents are

*author's email: valeria.riccardo@ccfe.ac.uk*

small. The Be tiles have much lower electrical resistivity, ~0.08µΩm (at 200 °C) and eddy currents become an issue. The currents interact with the local magnetic field to produce a torque, which has been estimated analytically or using finite element codes or both for most ILW Be tiles. Further discussion of the merits of each approach can be found in [2]. Calculation of the torque due to eddy current on single block Be tiles filling the space envelope of the original C-based tiles show that the total load on the existing supports is intolerable. The method chosen to reduce this load is to cut the Be tiles in slices, so decreasing the loop area and therefore the induced currents. An additional reduction of the eddy current is provided by the plasma facing surface being castellated, as discussed in further detail in Section 2.2), so locally resulting in an increased effective electrical resistivity.

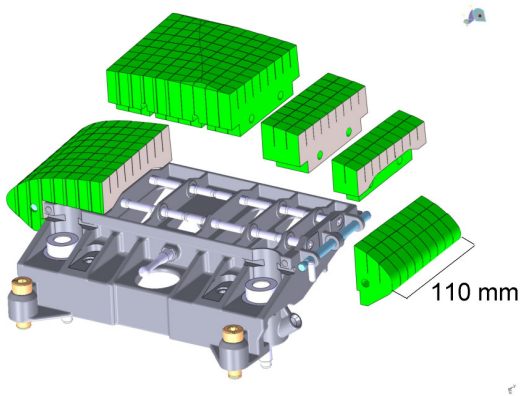

Fig.2 Typical Be tile assembly "exploded"

The slices require a carrier for support, which needs to withstand electromechanical forces from both eddy and halo current as well as allow the Be to thermally deform. Be and inconel have similar expansion coefficients, but they are not well coupled thermally and so temperature differences of ~200°C can occur and this demands suitable expansion clearances. The carrier design is driven by its interfaces with the Be slices, the existing support beam, the weight requirement for RH and the limited space available. The resulting complex geometry and the large number required of each type make casting the most appropriate manufacturing route. Vacuum-cast inconel 625 was selected as the best compromise between cost and material properties.

## 2.2 Compatibility with thermo-mechanical stresses

The need to avoid surface melting and to have an adequate fatigue life drives the design of the Be slice and its attachments.

Surface melting results in a thermal capability of 6MW/m$^2$ for 10s cold start (200°C) and 4MW/m$^2$ for 10s hot start (400°C).

The tile is subject to temperature, and hence stress, cycles. The aim is to minimize thermally induced stress to maximize the fatigue life. The thermal stresses depend on the temperature profile and the degree of constraint in the tiles. Global constraints have been eliminated by using kinematic supports which allow the tile to both expand and bow. At the surface, castellations have been introduced and sized following an extensive program of coupled thermal and mechanical analyses. With castellations, the surface cyclic strain is eliminated and the critical point becomes the castellation root. The optimum castellation size is 6mm×6mm square and 16mm deep with a keyhole profile as this gave similar thermal and mechanical performance, but for practical reasons the area has been increased to 12mm×12mm and the keyhole replaced by a half-round.

## 2.3 Optimization of power handling

The tile assemblies are separated by a finite gap (2–3 mm) with its lowest limit defined by the limiter curvature and space for RH tools and cumulative tolerances. The slices are castellated. Therefore edges exist not only from tile to tile but also within the tile itself. The power density on toroidally facing surfaces is 10 to100 times larger than on the plasma facing surface. Therefore edge exposure must be minimized, by shadowing, to avoid hot spot melting. To shadow an edge the surface upstream needs to be raised and is subject to higher power density. Consequently the less the edges need shadowing the better the overall power handling. The design limit for the exposed depth, ~40µm, is determined from sensitivity studies; this is combined with the minimum achievable cut width, 350µm, to give a limit for the slope of the central slice where edges are not shadowed. In addition, wherever possible, and definitely on all inner and outer limiters, no plasma facing holes are allowed and assembly needs to start with the bottom tile, with successive tiles covering the bolts of the tile below.

The plasma facing surface of inner and outer limiters is designed for the expected scrape-off layer decay length in limiter operation [3]. The surface is defined by polynomials equations and optimized to give constant power density on the tile surface. Where necessary, e.g. outer wall, to allow shadowing from tile to tile the poloidal direction is chamfered. The symmetry of the tile allows single field-line helicity. The analytical surfaces are translated in 3D models for manufacture, as well as numerical checking of the shadowing [4].

## 3. Manufacture of ILW Be PFCs

To minimize the cost of the project each step was managed by the project rather than by a main contractor. The material for the Be tiles was procured and free-issued to the Be manufacturers. Cast carriers were procured and free-issued for final machining. Individual components for the Be tile assemblies were delivered to JET and grouped into 'loose assemblies' before being sent to the Be manufacturer to be assembled and inspected by JET personnel, to minimize the handling of

Be at JET. As part of the final inspection, complete beams were assembled onto specially manufactured jigs that carefully represented the internal geometry of the JET machine and photogrametry surveys conducted to confirm the high accuracy requirements of the plasma facing surfaces.

### 3.1 Be re-cycling

JET had used significant quantities of Be in the past. The grade used was Brush Wellman S65-C VHP and this had become the baseline for the ITER. This choice was driven by the low iron content of S65-C leading to higher resistance to thermal shock. JET had ~4000kg of Be available for re-use. The procurement of the new material, a HIPed (hot iso-static pressed) version of S65-C cheaper to produce and of higher strength, could have been speeded up by recycling the JET stock, which helped achieving the target Fe concentration of 0.075%. However this had a tritium history and initially only 20kg were allowed at the supplier for tests. These tests satisfied the supplier's regulator, so the existing stock was recycled demonstrating that Be could be an asset rather than a liability when JET is eventually decommissioned.

### 3.2 Machining of castellations

Two EDM (electro discharge machining) rates were tested, 3.03cc/hr and 0.94cc/hr, on a number of samples later subjected to mechanical tests and were inspected with SEI (secondary electron images), BEI (backscattered electron images) and EDX (Energy Dispersive X-ray) photography. The tests results showed the two rates being statistically undistinguishable, so the faster rate was prescribed.

Seven 5x5 castellation samples were made with the chosen machining rate and tested to select the flash etch procedure removing the most EDM contaminants and modifying the least the castellation geometry. A 2 x 2 etching test matrix was performed using two different acid mixtures (de-ionized water with 1%HF-25%$HNO_3$ or 50%$HNO_3$) and two application methods (ultrasonic bath or peristaltic pump), with a fifth sample untreated and two samples left for further analysis. Before and after all tests performed, the width of the 16-ea grooves along the outer perimeter was measured using a non-contact video-optical comparator. Essentially no beryllium was removed with 50%$HNO_3$, ~11-μm were removed with ultrasonic agitation in 1%HF-25%$HNO_3$ and 0-4-μm were removed with the acid mixture flowing.

Later the samples were sectioned (see Fig.3) from the back to avoid disruption the area to be analyzed. Three areas were used for examination. Area1 was used to measure groove and corner radii dimension using a non-contact optical video comparator. Area2 was mounted in phenolic resin, ground and polished for metallographic examination of the cross section of the castellation and corner radii measurement. Area3 was used for surface roughness measurement and SEM/EDX examination including corner radii measurement. All grooves were narrower than those at the perimeter. The sharpest average radius was in the unetched sample at 0.027-mm. The next sharpest corners were with flowing acid by 1%HF-25%$HNO_3$ (0.029-mm) followed by 50%$HNO_3$ (0.030-mm). The corners were significantly less sharp for samples subjected to ultrasonically agitated etch. In all EDX images Cu and Zn were detected. Both samples etched with 1%HF-25%$HNO_3$ showed less distinct Cu and Zn peaks, with substantially higher background which is typically seen in "clean" areas. Because of the chemical and geometrical benefits the 1%HF-25%$HNO_3$ solution was selected.

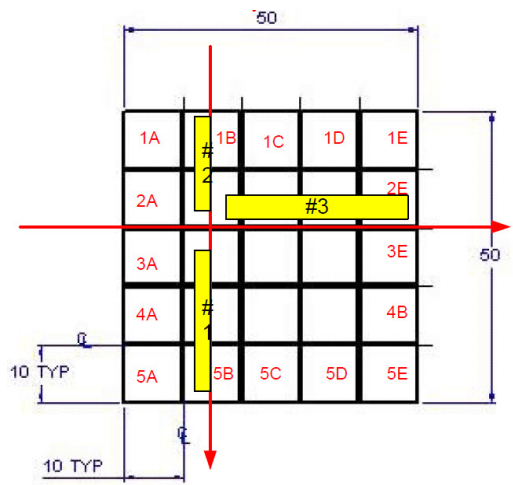

Fig.3 Castellation flash etch sample

### 3.3 Prototypes and jigs

Preproduction prototypes of the most complex tile assemblies were made as part of the design validation. Cast carriers were made to confirm the required material properties could be achieved. Cast carriers were machined to confirm they could be machined to the required tolerances. Aluminum slices were machined and assembled to the pre-prototype carriers to identify any design and production issues and validate the design of the tile assemblies. In addition, the Be machinists had to qualify their EDM and etch process to confirm they met those selected in the pre-production phase.

As all Be tiles were planned and designed for RH installation, substantial effort was spent to design and manufacture RH tools for the new tiles [5]. To give RH installation the greatest chance of success, a number of jigs and mock-ups (simplified versions of the real components representative in terms of physical size, weight, and centre of gravity with correct interfaces to both tools and neighboring components) were also designed and manufactured. Jigs and mock-ups allowed the test-fitting of assemblies on to replicas of the interior of the machine and trial fitting of prototypes using the same RH techniques as in the final installation. The necessary installation procedures were developed in conjunction with the mock-up trials.

## 4. Installation of ILW Be PFCs and other shutdown issues

As part of the tile assembly acceptance process, these were installed on jigs and subjected to a gap gun inspection to measure the gap and step between tiles assemblies. The gap gun inspection was repeated once the tile assemblies had been installed as shown in Fig.4 to confirm they were located correctly and it several occasions it was useful to highlight potential issues.

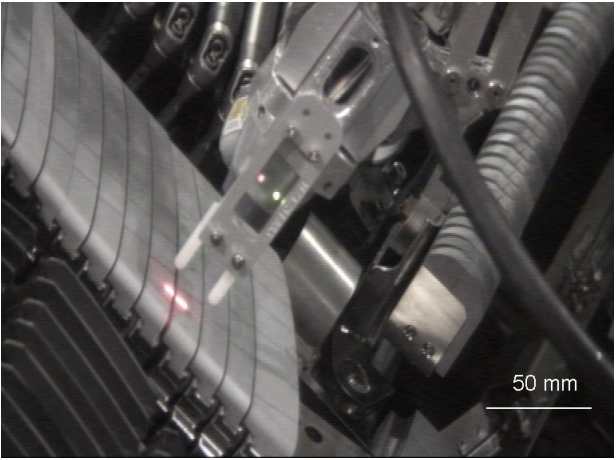

Fig.4 Inspection of the installed Lower Hybrid frame

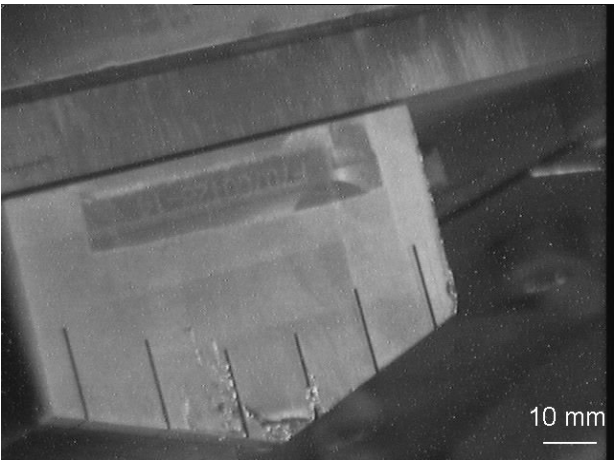

Fig.5 Disruption damage to the upper dump plate tiles

Despite all the risk reduction measures implemented to maximize the shutdown efficiency, a large number of anomalies (for example clashes of the new components with existing components not fully captured by the configuration model) was encountered and had to be resolved, more often than not by modifying the new components. Typical examples are (1) components being fully removed from the configuration file, while their supports (or part of these) were still in vessel and could not be removed; (2) components being of a slightly larger size than recorded; (3) supports not being at their nominal location or angle; (4) uncertainties in the physical gap between components separately surveyed. In total 227 anomalies have beer recorded during the ILW shutdown, of these 159 resulted in non-conformance reports (NCR), and of these 33 in major modifications or new projects.

## 5. Engineering aspects operation with ILW Be PFCs

### 5.1 Conditioning

Restarting the plasma on the ITER-like wall proved much easier than for carbon with no further conditioning required after the first plasma and no deconditioning by disruptions [6].

### 5.2 Disruptions and off-normal events

Disruptions have caused the melting of the outermost upper dump plate tiles (Fig.5) due to field line penetration through the large gaps between protections on the outer wall. The reason why disruptions have been so threatening to the ILW is also due to the reduced radiative losses compared to carbon [7], leaving most of the plasma magnetic energy to be conducted to the wall and revealing that as with ITER, disruption avoidance and mitigation strategies are needed routinely.

Slide-away electron beams due to lack of fuelling following an emergency coil shutdown immediately after a successful breakdown caused the melting of 1-2 castellations (fig. 6) in a region normally far from the separatrix on the inner wall. This occurred within 100 pulses from first plasma and highlighted a weakness in the plant response which was not exercised by plasmas on the C-wall (as they would not survive such a breakdown) and was promptly addressed. The damaged caused by this event was monitored during plasmas and in fortnightly inspections; it only marginally increased over the operating period and had no effect on plasmas, so it will be left untouched.

### 5.3 Power handling

The main chamber power handling has achieved and possibly exceeded the design targets.

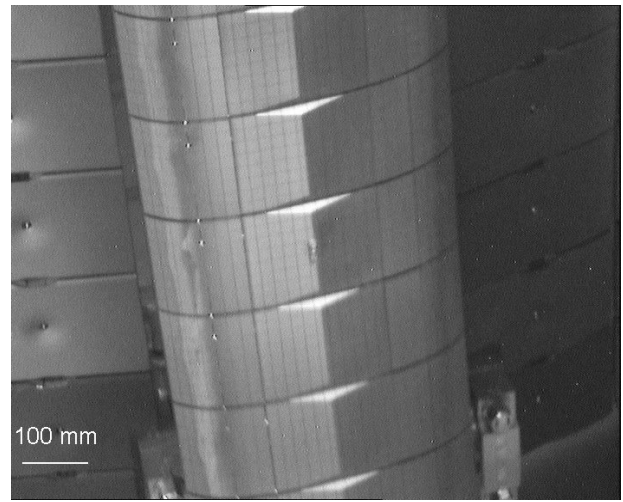

Fig.6 Slide-away damage on inner wall guard limiter

To test the power handling limits and validate the assumptions of the design of the Be tile assemblies, experiments have been performed with varying additional power both in limiter and in X-point configuration, in L-mode as well as in H-mode. Two of the design inner limiter and two of the design outer limiter configurations have been run and compared with the numerical results. Although the configurations had to be slightly modified to be run, the measured and the computed power footprints are very similar [8]. For the smaller elongation limiter configuration the predicted performance degradation caused by the decision not to shadow poloidally facing edges of the inner wall limiter has been confirmed by the experiments. While calculations predict a 20-30% contribution to the temperature increase due to edge heat flux, experiments show ~50% additional increase in temperature near the exposed edge.

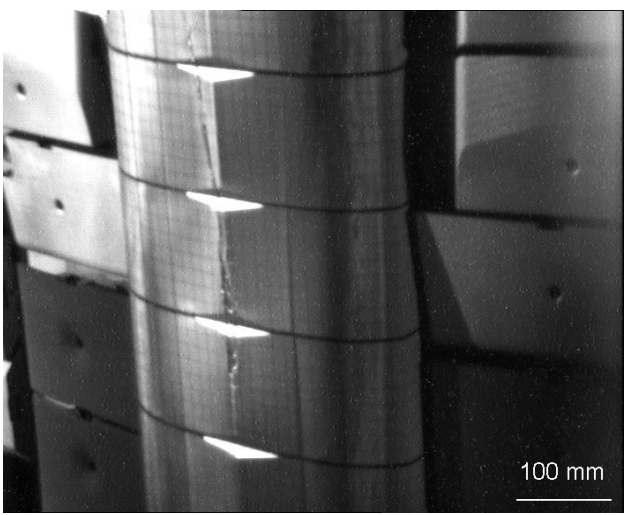

Fig.7 Thermal damage on inner wall guard limiter

The higher elongation limiter configuration was run with up to 5MW additional heating and 1MW ohmic heating for over 7s on the inner wall. The monitored temperature stayed <760°C; however, the temperature on beams outside the monitored regions of interest reached far higher temperatures, with one in view of a camera apparently releasing Be in bursts. On inspection a number of beams showed various extent of melting damage (Fig.7 shows the worst location). As the damage occurred in a very unusual mode of operation and has not affected operation since (albeit only two weeks with a single plasma type featuring a very prompt X-point formation), these tiles will be monitored, not replaced.

## 6. Conclusions

The design of the Be tile assemblies was the result of a complex compromise to satisfy requirements and technical boundary conditions. Key to the practical implementation of the design was the attention given to detail during manufacturing: pre-production prototypes, jigs and inspections aimed to minimize risk of in-vessel clashes. Several errors in the configuration control of the existing components required tile assemblies to be modified. In some cases uncertainty in the configuration was known and insufficient per-installation surveys were carried out; for any future major modification adequate surveys must be considered.

Despite the commissioning of the protection systems continued for most of the operational period, the Be tiles survived and delivered better than expected. However, off-normal (slide-away and disruptions) events and prolonged heated limiter configurations caused some damage.

## Acknowledgments

This work was funded by EURATOM via the JET Operation Contract within the framework of EFDA. The views and opinions expressed herein do not necessarily reflect those of the European Commission.